\RequirePackage{lineno}
\documentclass[aps,prl,showpacs,letterpaper,twocolumn,longbibliography,superscriptaddress]{revtex4-1}%
\usepackage{graphicx}
\usepackage{bm}
\usepackage{epsf}
\usepackage{rotating}
\usepackage{epsfig,graphics,rotate,color}
\usepackage{wrapfig}
\usepackage{amssymb}
\usepackage{amsmath}
\usepackage{amsfonts}
\usepackage{array,hhline,dcolumn}%
\usepackage{gensymb}
\usepackage{placeins}
\usepackage{tabularx}
\usepackage[colorlinks=true,citecolor=blue,linkcolor=blue]{hyperref}
\usepackage{multirow, makecell}
\usepackage{booktabs}
\usepackage{siunitx}
\usepackage{adjustbox}

\usepackage[utf8]{inputenc}

\usepackage{physics}
\newcommand\headercell[1]{%
   \smash[b]{\begin{tabular}[t]{@{}c@{}} #1 \end{tabular}}}
\usepackage{accents}
\usepackage{color}
\usepackage{xcolor}
\usepackage[normalem]{ulem}

\begin{document}


\title[Article Title]{Searching for Decoherence from Quantum Gravity at the IceCube South Pole Neutrino Observatory}

\affiliation{III. Physikalisches Institut, RWTH Aachen University, D-52056 Aachen, Germany}
\affiliation{Department of Physics, University of Adelaide, Adelaide, 5005, Australia}
\affiliation{Dept. of Physics and Astronomy, University of Alaska Anchorage, 3211 Providence Dr., Anchorage, AK 99508, USA}
\affiliation{Dept. of Physics, University of Texas at Arlington, 502 Yates St., Science Hall Rm 108, Box 19059, Arlington, TX 76019, USA}
\affiliation{CTSPS, Clark-Atlanta University, Atlanta, GA 30314, USA}
\affiliation{School of Physics and Center for Relativistic Astrophysics, Georgia Institute of Technology, Atlanta, GA 30332, USA}
\affiliation{Dept. of Physics, Southern University, Baton Rouge, LA 70813, USA}
\affiliation{Dept. of Physics, University of California, Berkeley, CA 94720, USA}
\affiliation{Lawrence Berkeley National Laboratory, Berkeley, CA 94720, USA}
\affiliation{Institut f{\"u}r Physik, Humboldt-Universit{\"a}t zu Berlin, D-12489 Berlin, Germany}
\affiliation{Fakult{\"a}t f{\"u}r Physik {\&} Astronomie, Ruhr-Universit{\"a}t Bochum, D-44780 Bochum, Germany}
\affiliation{Universit{\'e} Libre de Bruxelles, Science Faculty CP230, B-1050 Brussels, Belgium}
\affiliation{Vrije Universiteit Brussel (VUB), Dienst ELEM, B-1050 Brussels, Belgium}
\affiliation{Department of Physics and Laboratory for Particle Physics and Cosmology, Harvard University, Cambridge, MA 02138, USA}
\affiliation{Dept. of Physics, Massachusetts Institute of Technology, Cambridge, MA 02139, USA}
\affiliation{Dept. of Physics and The International Center for Hadron Astrophysics, Chiba University, Chiba 263-8522, Japan}
\affiliation{Department of Physics, Loyola University Chicago, Chicago, IL 60660, USA}
\affiliation{Dept. of Physics and Astronomy, University of Canterbury, Private Bag 4800, Christchurch, New Zealand}
\affiliation{Dept. of Physics, University of Maryland, College Park, MD 20742, USA}
\affiliation{Dept. of Astronomy, Ohio State University, Columbus, OH 43210, USA}
\affiliation{Dept. of Physics and Center for Cosmology and Astro-Particle Physics, Ohio State University, Columbus, OH 43210, USA}
\affiliation{Niels Bohr Institute, University of Copenhagen, DK-2100 Copenhagen, Denmark}
\affiliation{Dept. of Physics, TU Dortmund University, D-44221 Dortmund, Germany}
\affiliation{Dept. of Physics and Astronomy, Michigan State University, East Lansing, MI 48824, USA}
\affiliation{Dept. of Physics, University of Alberta, Edmonton, Alberta, Canada T6G 2E1}
\affiliation{Erlangen Centre for Astroparticle Physics, Friedrich-Alexander-Universit{\"a}t Erlangen-N{\"u}rnberg, D-91058 Erlangen, Germany}
\affiliation{Physik-department, Technische Universit{\"a}t M{\"u}nchen, D-85748 Garching, Germany}
\affiliation{D{\'e}partement de physique nucl{\'e}aire et corpusculaire, Universit{\'e} de Gen{\`e}ve, CH-1211 Gen{\`e}ve, Switzerland}
\affiliation{Dept. of Physics and Astronomy, University of Gent, B-9000 Gent, Belgium}
\affiliation{Dept. of Physics and Astronomy, University of California, Irvine, CA 92697, USA}
\affiliation{Karlsruhe Institute of Technology, Institute for Astroparticle Physics, D-76021 Karlsruhe, Germany }
\affiliation{Karlsruhe Institute of Technology, Institute of Experimental Particle Physics, D-76021 Karlsruhe, Germany }
\affiliation{Dept. of Physics, Engineering Physics, and Astronomy, Queen's University, Kingston, ON K7L 3N6, Canada}
\affiliation{Department of Physics {\&} Astronomy, University of Nevada, Las Vegas, NV, 89154, USA}
\affiliation{Nevada Center for Astrophysics, University of Nevada, Las Vegas, NV 89154, USA}
\affiliation{Dept. of Physics and Astronomy, University of Kansas, Lawrence, KS 66045, USA}
\affiliation{Centre for Cosmology, Particle Physics and Phenomenology - CP3, Universit{\'e} catholique de Louvain, Louvain-la-Neuve, Belgium}
\affiliation{Department of Physics, Mercer University, Macon, GA 31207-0001, USA}
\affiliation{Dept. of Astronomy, University of Wisconsin{\textendash}Madison, Madison, WI 53706, USA}
\affiliation{Dept. of Physics and Wisconsin IceCube Particle Astrophysics Center, University of Wisconsin{\textendash}Madison, Madison, WI 53706, USA}
\affiliation{Institute of Physics, University of Mainz, Staudinger Weg 7, D-55099 Mainz, Germany}
\affiliation{Department of Physics, Marquette University, Milwaukee, WI, 53201, USA}
\affiliation{Institut f{\"u}r Kernphysik, Westf{\"a}lische Wilhelms-Universit{\"a}t M{\"u}nster, D-48149 M{\"u}nster, Germany}
\affiliation{Bartol Research Institute and Dept. of Physics and Astronomy, University of Delaware, Newark, DE 19716, USA}
\affiliation{Dept. of Physics, Yale University, New Haven, CT 06520, USA}
\affiliation{Columbia Astrophysics and Nevis Laboratories, Columbia University, New York, NY 10027, USA}
\affiliation{Dept. of Physics, University of Oxford, Parks Road, Oxford OX1 3PU, United Kingdom}
\affiliation{Dipartimento di Fisica e Astronomia Galileo Galilei, Universit{\`a} Degli Studi di Padova, 35122 Padova PD, Italy}
\affiliation{Dept. of Physics, Drexel University, 3141 Chestnut Street, Philadelphia, PA 19104, USA}
\affiliation{Physics Department, South Dakota School of Mines and Technology, Rapid City, SD 57701, USA}
\affiliation{Dept. of Physics, University of Wisconsin, River Falls, WI 54022, USA}
\affiliation{Dept. of Physics and Astronomy, University of Rochester, Rochester, NY 14627, USA}
\affiliation{Department of Physics and Astronomy, University of Utah, Salt Lake City, UT 84112, USA}
\affiliation{Oskar Klein Centre and Dept. of Physics, Stockholm University, SE-10691 Stockholm, Sweden}
\affiliation{Dept. of Physics and Astronomy, Stony Brook University, Stony Brook, NY 11794-3800, USA}
\affiliation{Dept. of Physics, Sungkyunkwan University, Suwon 16419, Korea}
\affiliation{Institute of Physics, Academia Sinica, Taipei, 11529, Taiwan}
\affiliation{Dept. of Physics and Astronomy, University of Alabama, Tuscaloosa, AL 35487, USA}
\affiliation{Dept. of Astronomy and Astrophysics, Pennsylvania State University, University Park, PA 16802, USA}
\affiliation{Dept. of Physics, Pennsylvania State University, University Park, PA 16802, USA}
\affiliation{Dept. of Physics and Astronomy, Uppsala University, Box 516, S-75120 Uppsala, Sweden}
\affiliation{Dept. of Physics, University of Wuppertal, D-42119 Wuppertal, Germany}
\affiliation{Deutsches Elektronen-Synchrotron DESY, Platanenallee 6, 15738 Zeuthen, Germany }

\author{R. Abbasi}
\affiliation{Department of Physics, Loyola University Chicago, Chicago, IL 60660, USA}
\author{M. Ackermann}
\affiliation{Deutsches Elektronen-Synchrotron DESY, Platanenallee 6, 15738 Zeuthen, Germany }
\author{J. Adams}
\affiliation{Dept. of Physics and Astronomy, University of Canterbury, Private Bag 4800, Christchurch, New Zealand}
\author{S. K. Agarwalla}
\thanks{also at Institute of Physics, Sachivalaya Marg, Sainik School Post, Bhubaneswar 751005, India}
\affiliation{Dept. of Physics and Wisconsin IceCube Particle Astrophysics Center, University of Wisconsin{\textendash}Madison, Madison, WI 53706, USA}
\author{J. A. Aguilar}
\affiliation{Universit{\'e} Libre de Bruxelles, Science Faculty CP230, B-1050 Brussels, Belgium}
\author{M. Ahlers}
\affiliation{Niels Bohr Institute, University of Copenhagen, DK-2100 Copenhagen, Denmark}
\author{J.M. Alameddine}
\affiliation{Dept. of Physics, TU Dortmund University, D-44221 Dortmund, Germany}
\author{N. M. Amin}
\affiliation{Bartol Research Institute and Dept. of Physics and Astronomy, University of Delaware, Newark, DE 19716, USA}
\author{K. Andeen}
\affiliation{Department of Physics, Marquette University, Milwaukee, WI, 53201, USA}
\author{G. Anton}
\affiliation{Erlangen Centre for Astroparticle Physics, Friedrich-Alexander-Universit{\"a}t Erlangen-N{\"u}rnberg, D-91058 Erlangen, Germany}
% ...
\date{\today}

\collaboration{IceCube Collaboration}
\noaffiliation


\begin{abstract}Neutrino oscillations at the highest energies and longest baselines provide a natural quantum interferometer with which to study the structure of spacetime and test the fundamental principles of quantum mechanics.  If the metric of spacetime has a quantum mechanical description, there is a generic expectation that its fluctuations at the Planck scale would introduce non-unitary effects that are inconsistent with the standard unitary time evolution of quantum mechanics.  Neutrinos interacting with such fluctuations would lose their quantum coherence, deviating from the expected oscillatory flavor composition at long distances and high energies.  The IceCube South Pole Neutrino Observatory is a billion-ton neutrino telescope situated in the deep ice of the Antarctic glacier. Atmospheric neutrinos detected by IceCube in the energy range 0.5--10 TeV have been used to test for coherence loss in neutrino propagation. No evidence of anomalous neutrino decoherence was observed, leading to the strongest experimental limits on neutrino-quantum gravity interactions to date, significantly surpassing expectations from natural Planck-scale models.  The resulting constraint on the effective decoherence strength parameter within an energy-independent decoherence model is $\Gamma_0\leq 1.17\times10^{-15}$~eV, improving upon past limits by a factor of 30.  For decoherence effects scaling as E$^2$, limits are advanced by more than six orders of magnitude beyond past measurements.  
\end{abstract}


\maketitle

The construction of a consistent and predictive quantum theory of gravity is an outstanding challenge in fundamental physics.  A central experimental and theoretical question is whether the metric of spacetime exhibits the quantum fluctuations that are intrinsic to all other known fundamental fields.  Such fluctuations would represent quantum perturbations in the geometry of spacetime itself, most pronounced on Planck-scale distances or times.  At scales below the Planck energy $E_P$ this sea of  ``spacetime foam''~\cite{wheeler_fluct} could induce small modifications to standard quantum mechanical time evolution rule of propagating particles, leading to non-unitary effects~\cite{carlip2023spacetime}.  Testing for these small violations is one of the few clear experimental avenues through which searches for quantum gravity can be conducted at the single particle level~\cite{carney2019tabletop}.

The oscillations of massive neutrinos between flavor states is a quantum process that has been widely studied over many energies and baselines and with a multitude of neutrino production and detection techniques~\cite{hagiwara2002review}.  Because they interact only through the weak force and gravity, neutrinos are largely isolated from their surroundings and rarely interact as they propagate through matter.  This isolation allows quantum coherence to be exhibited over distance scales of thousands of kilometers, enabling neutrino oscillations to serve as a precise interferometer for fundamental studies of the quantum nature of spacetime.  

\begin{figure}[t]
\centering
\includegraphics[width=.48\textwidth]{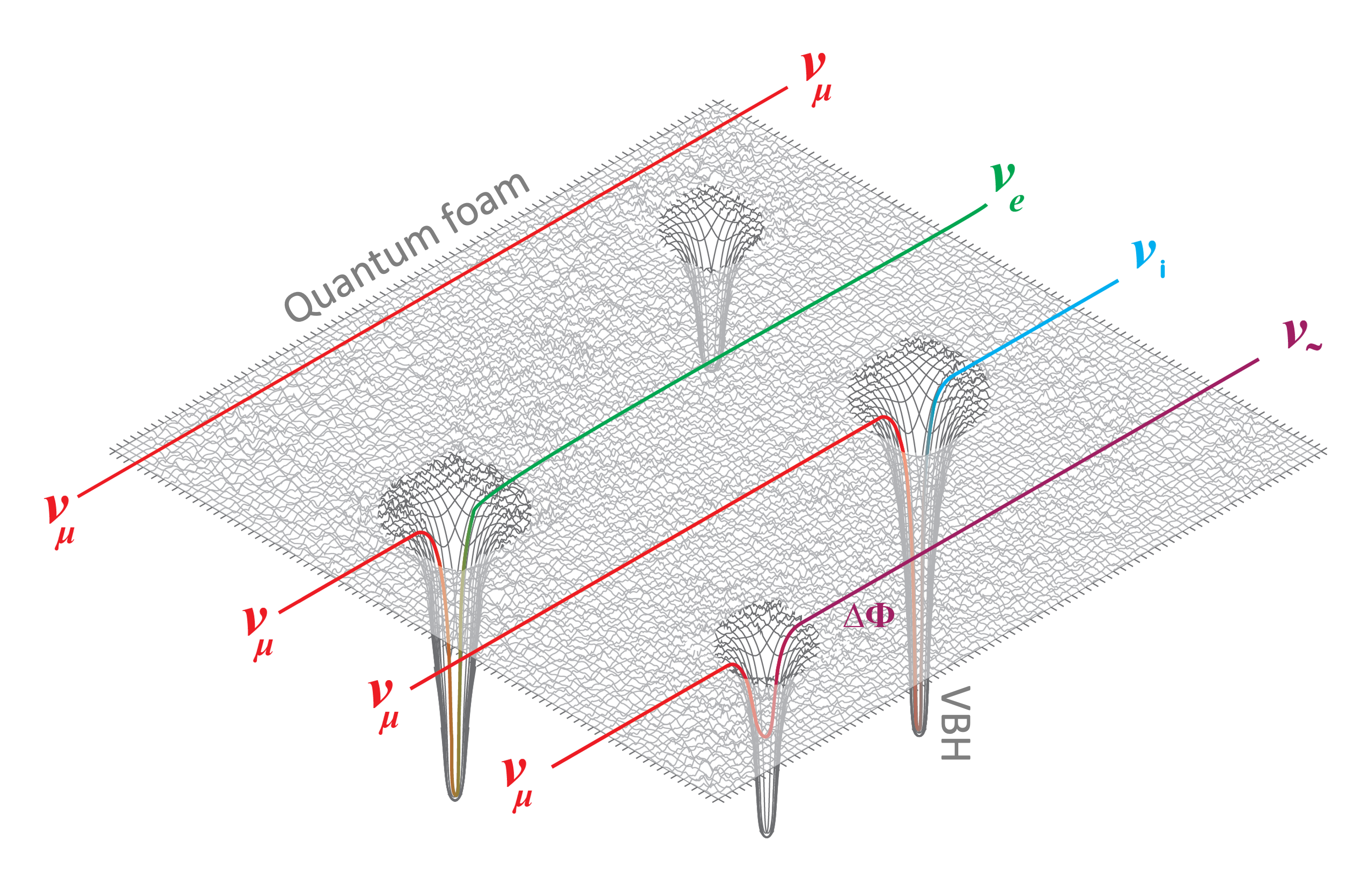}
\caption{\textit{\textbf{Illustration of the quantum gravitational decoherence effect.}} Interactions of neutrinos with fluctuating spacetime leads to decoherence of neutrino oscillations through non-unitary time evolution. Roman indices represent mass states, $\nu_\mu$ and $\nu_e$ are flavor states, and $\nu_\sim$ a phase-perturbed state carrying neither definite mass or flavor.}
\label{fig:jacksketch}
\end{figure}

Oscillations of neutrinos produced in cosmic-ray air showers (termed atmospheric neutrinos) have been experimentally verified to maintain quantum coherence over distance scales of at least the diameter of the Earth (1.2$\times$10$^4$ km)~\cite{jung2001oscillations,gaisser2002flux}.  If propagating neutrinos were to exchange quantum information about their flavor or mass with a fluctuating environment, or to experience stochastic perturbations to their quantum phases, their coherence would be lost during travel~\cite{schlosshauer2007decoherence}.  Coherence loss causes distinct initial state wave functions to produce equivalent final states, in violation of quantum mechanical unitarity. Its observation would therefore be a smoking-gun signature of neutrinos undergoing quantum gravitational effects that would be difficult to explain by other means~\cite{carney2019tabletop}.

\begin{figure}[t]
\centering
\includegraphics[width=.48\textwidth]{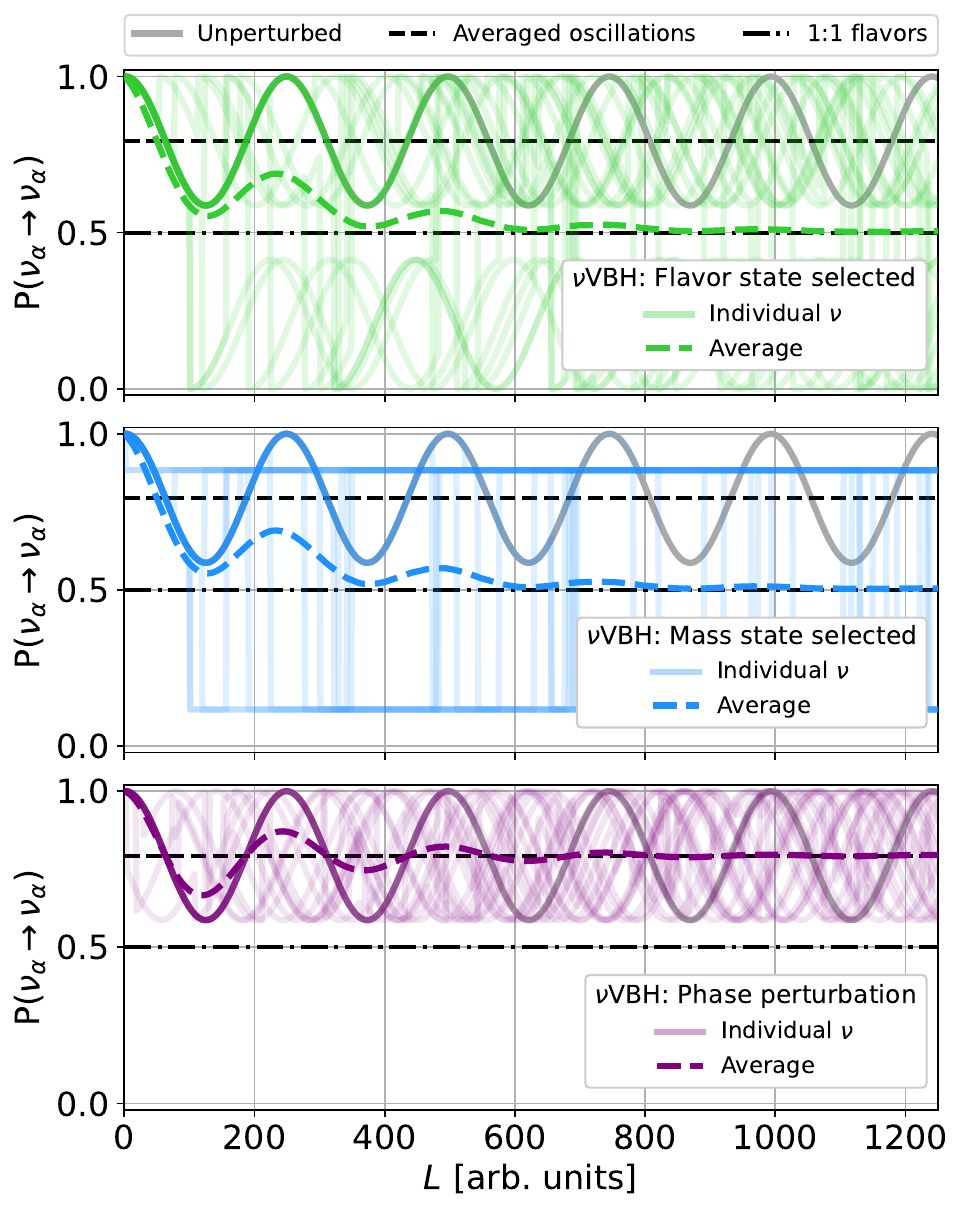}
\caption{\textit{\textbf{Decoherence of an oscillating neutrino ensemble.}} Non-unitary oscillation behaviour can emerge from various types of interactions between neutrinos and VBHs, including absorption with emission in a random flavor eigenstate (top row), absorption of a neutrino with emission in a random mass eigenstate (middle row), or a random perturbation to the neutrino phase (bottom row).  The full phenomenology of these models is described in Ref.~\cite{stuttard2020neutrino}. }
\label{fig:fig1}
\end{figure}

The signatures of neutrino decoherence include both a damping of neutrino flavor transitions at large distances, and a non-unitary flavor evolution below the oscillation wavelength, which is longer at the higher energies.  A general description is provided by the formalism of open quantum systems whereby new superoperators are inserted into the Lindblad master equation~\cite{schlosshauer2007decoherence}. We follow the original proposal of Hawking~\cite{hawking_vbh,vbh_theory} and investigate a class of theories where the mechanism for coherence loss is the interaction between propagating particles and virtual black holes (VBHs)  generated in quantum fluctuations of the spacetime metric, as shown schematically in Fig.~\ref{fig:jacksketch}.  When a neutrino is absorbed and re-radiated with its wave function either collapsed or its phase perturbed~\cite{stuttard2020neutrino}, a stochastic contribution to its time evolution is introduced leading to coherence loss (Fig.~\ref{fig:fig1}).  Although relying on a different underlying mechanism, $\nu$VBH interaction models of decoherence exhibit strikingly similar phenomenology to theories advanced by Penrose~\cite{penrose1996gravity}, and Diosi~\cite{diosi2014gravitation}, whereby fluctuations in the metric encode an intrinsic quantum uncertainty into the phase of particles propagating in otherwise empty spacetime through uncertainty introduced into the time coordinate or path length. Because energy is the cause of spacetime curvature, and also because the effects of Planck-scale theories are expected to be suppressed at lower energy scales, tests of decoherence benefit from using the highest possible energies, while measurements over the longest possible baselines allow even miniscule effects to accumulate into potentially measurable signals. 

The IceCube Neutrino Observatory~\cite{aartsen2017icecube}, a neutrino detector within the glacial ice of the Geographic South Pole, occupies 1 $\text{km}^3$ of ice from 1450-2450 m under the surface. A total of 5,160 Digital Optical Modules (DOMs) \cite{icdaq} are distributed among 86 cables, with a more-densely instrumented sub-array called DeepCore \cite{deepcoredesign} located at the center of the detector.  IceCube has detected two major populations of high energy neutrinos: astrophysical neutrinos~\cite{aartsen2014observation} which dominate the flux of $\nu_\mu$ above 200~TeV and traverse cosmological baselines; and atmospheric neutrinos with a rate peaking around 1~TeV in IceCube, and baselines of up to the Earth diameter.  Although they have the longest baselines and highest energies, searches for decoherence with astrophysical neutrinos are limited by not knowing the oscillation baseline to within an oscillation wavelength, and the unknown flavor composition at the source. While these neutrinos may be used to test for other violations of Lorentz-symmetric quantum mechanical time evolution~\cite{icecube2018neutrino}, the above considerations prohibit their unambiguous use for decoherence searches. The large ensemble of high-energy atmospheric neutrinos detected by IceCube, however, present an especially compelling window through which to seek evidence of quantum gravitational effects at the single-particle level.  

Charged-current (CC) interactions of neutrinos with matter produce charged leptons that emit Cherenkov light in identifiable distributions within the detector, which can be categorized into two basic morphologies: tracks and cascades. The electrons from $\nu_e$ CC interactions yield electromagnetic showers with a roughly spherical distribution of photons (cascade), whereas muons from $\nu_\mu$ CC interactions emit light along a linear trajectory (track).  Track events can be further sub-divided into two categories:  ``starting tracks'' emanating from the detector volume itself, and ``through-going tracks'' where the neutrino interacted in the ice or rock beneath. The latter category is statistically dominant in this energy range, due to the much larger effective target volume. Tau leptons from $\nu_\tau$ CC interactions have signals with characteristics of both track and cascade morphologies. In addition, neutrinos of all flavors can have neutral-current (NC) interactions with matter that produce hadronic showers, in turn inducing cascade signals. This analysis focuses on through-going tracks, which provide a large, high purity sample of events with which to search muon-neutrino disappearance signatures induced by decoherence.   

Two forms of $\nu$VBH interaction have been tested for using the energy and zenith spectrum of these tracks: neutrino mass state phase perturbation (hereafter \textit{phase perturbation}) and democratic mass/flavor state selection (hereafter \textit{state selection}).   These specific flavor models have been selected for concreteness, but the limits obtained are similar for each and our results can be interpreted broadly as constraints on the scale of decoherence of neutrino oscillations induced by coupling to spacetime foam.

\section{Decoherence in Neutrino Oscillations }

Calculating the flavor oscillations of neutrinos detected by IceCube requires consideration of a plethora of effects, including vacuum oscillations~\cite{giunti2007fundamentals}, coherent forward-scattering from matter~\cite{Mikheev:1987qk}, neutrino absorption~\cite{cooper2011high,icecube2017measurement}, and $\tau$ regeneration~\cite{beacom2002enhanced,arguelles2022tau}. Because the ensemble of neutrinos is subject to both unitary and non-unitary effects such as absorption and re-interaction, the problem must be approached through the master equation formalism. We perform this calculation using the {\tt NuSquids} software package~\cite{arguelles2022nusquids}, by adding decoherence terms to the neutrino oscillation master equation governing time evolution of the neutrino reduced density matrix $\rho(t)$.  Our parameterization is explained in detail in Ref. ~\cite{stuttard2020neutrino} and briefly reviewed below.

\begin{figure}[t]
\centering
\includegraphics[width=.48\textwidth]{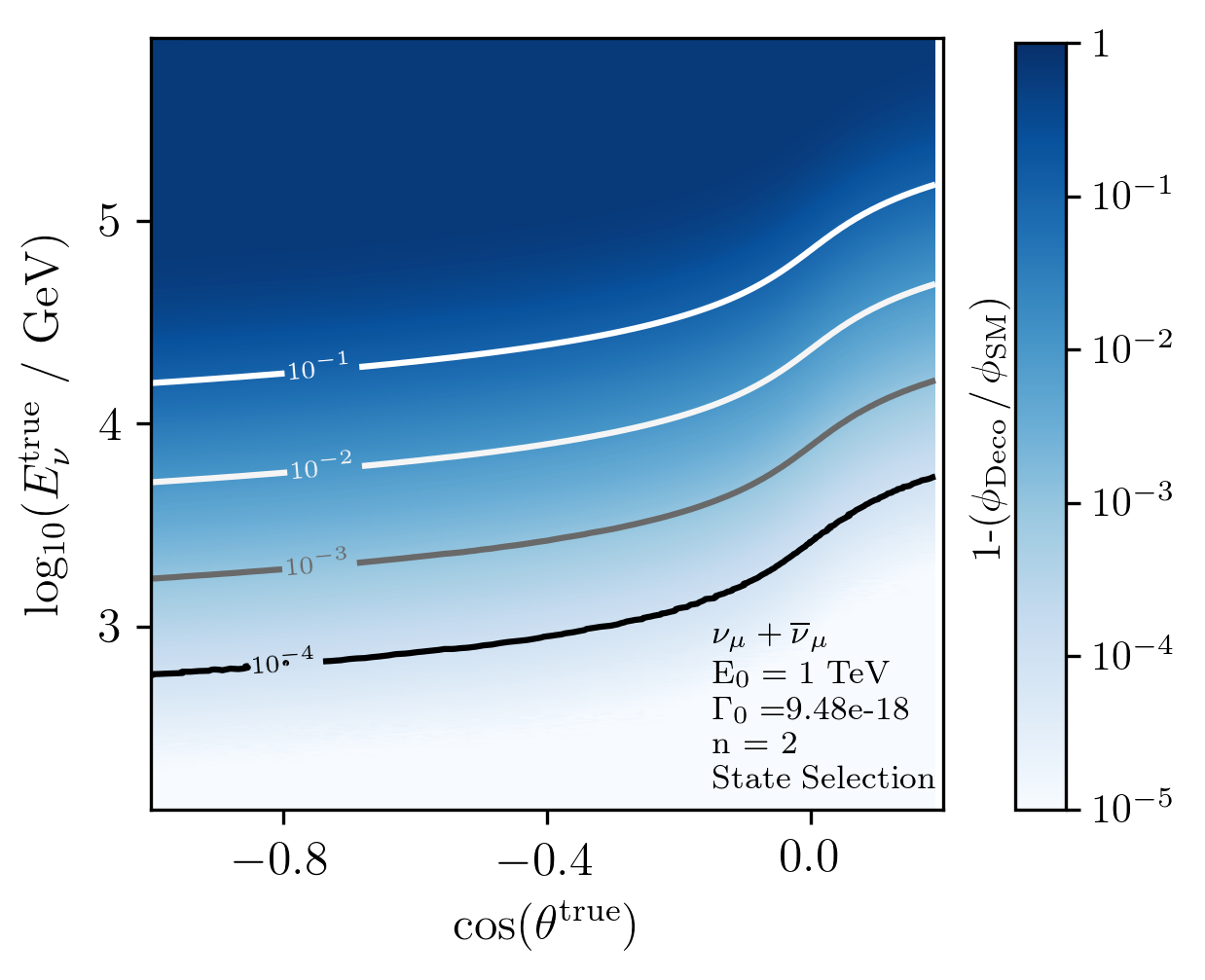}
\caption{\textit{\textbf{Muon neutrino oscillogram.}} Ratio of the predicted $\nu_\mu$ flux for the state selection model with the decoherence parameters listed versus the standard model prediction.}
\label{fig:fig2}
\end{figure}

Atmospheric neutrinos are produced predominantly through decays of charged pions and kaons~\cite{Fedynitch:2015zma}.  Production of $\nu_\tau / \bar{\nu}_\tau$ is highly suppressed, inhibited by the large mass of the $\tau^\pm$, lepton which can only be created through decays of heavy hadrons~\cite{bhattacharya2015perturbative}. Decoherence effects introduce muon neutrino disappearance ($\nu_\mu\rightarrow\nu_e$ and $\nu_\mu\rightarrow\nu_\tau$), and also increased fluxes of $\nu_\tau$ at all energies that exhibit a complex oscillation phenomenology.  High-energy $\nu_\tau$ undergoing CC interactions produce a $\tau^\pm$ lepton, which  decays weakly to a lower energy $\nu_\tau$ as well as producing secondary neutrinos of all flavors, including additional $\nu_\mu$~\cite{arguelles2022tau,safa2022taurunner}.  These secondary neutrinos also oscillate, and in the case of the secondary $\nu_\tau$ lead to further regeneration.  Because the neutrino flux is a steeply falling function of energy, the disappearance of $\nu_\mu$ through decoherence is a significantly larger effect than the appearance of $\nu_\mu$ through re-generation from higher-energy $\nu_\tau$, though both are included in our calculations.  An example oscillogram for a representative set of parameters, showing the change in $\nu_\mu$ flux as a function of energy and zenith angle across the high-energy IceCube $\nu_\mu$ sample is shown in Fig.~\ref{fig:fig2}.

The evolution of a neutrino system with Hamiltonian $H$ and decoherence superoperator $\mathcal{D}\left[ \rho \right]$ is described by (in natural units with $\hbar=c=1$),
\begin{linenomath*}
\begin{equation}
\label{lind_eq}
 \dot{\rho} = -i[H,\rho] - \mathcal{D}\left[ \rho \right].
\end{equation} 
\end{linenomath*}
The first term encodes the standard unitary time evolution that drives neutrino oscillations, and the second encapsulates the potentially non-unitary contributions that may be introduced through quantum gravitational effects. A convenient, general form of $\mathcal{D}\left[ \rho \right]$ is an expansion in the SU(3) basis \cite{Gago_deco,Benatti_open_quantum_system,operator_expansion_deco,Buoninfante_deco,stuttard2020neutrino} where
\begin{linenomath*}
\begin{equation}
\label{su3_exp}
 \mathcal{D}(\rho) = (D_{\mu \nu}\rho^\nu) b^\mu ,
\end{equation}
\end{linenomath*}
\noindent with $\rho^\nu$ as the density matrix projection along SU(3) basis vector $b^\mu$ (the Gell-Mann matrices). $D_{\mu \nu}$ is a $9\times 9$ matrix that parameterizes the decoherence effects on the neutrino system.
In the phase perturbation model (Fig.~\ref{fig:fig1}, bottom), the outgoing neutrino state emerges with one or two of the phases of the mass basis states distinctly perturbed. The effect on the average oscillation probability corresponds to a damping that follows
\begin{linenomath*}
\begin{equation}
D_{\mathrm{phase \, perturbation}} = \mathrm{diag}(0,\Gamma,\Gamma,0,\Gamma,\Gamma,\Gamma,\Gamma,0) ,
\end{equation}
\end{linenomath*}
\noindent where $\Gamma$ is the decoherence parameter, with dimensions of energy. At distances long relative to $1/\Gamma$, this model predicts a flux that tends towards an incoherent sum of mass eigenstates. 

The loss of neutrino flavor or mass information in a $\nu$VBH interaction is expected on the basis of non-conservation of global quantum numbers by black holes~\cite{Anchordoqui:2005gj}. This motivates models whereby $\nu$ emerging from VBH interactions are emitted in states collapsed randomly into a given mass or flavor basis state. State selection in either the mass or flavor bases (Fig.~\ref{fig:fig1}, top and middle) impose equivalent overall damping effects, leading to a flux equally weighted in all neutrino flavors at distances longer than 1/$\Gamma$. This democratization is independent of the initial flux and basis of the randomization~\cite{stuttard2020neutrino}.  In the SU(3) basis, the state selection model takes the form
\begin{linenomath*}
\begin{gather}
D_{\mathrm{state \, selection}} = \mathrm{diag}(0,\Gamma,\Gamma,\Gamma,\Gamma,\Gamma,\Gamma,\Gamma,\Gamma).
\end{gather}
\end{linenomath*}
Compared to the phase perturbation case, the state selection $D$ matrix has two additional non-zero terms. These are the third and eighth diagonal elements, sometimes referred to as the \textit{relaxation} terms~\cite{GUZZO2016408, Gomes:2020muc}.

In both models, the neutrino energy $E_\nu$ dependence of the damping factor $\Gamma$ is unknown. We thus test a representative set of models for $\Gamma(E_\nu)$, parameterized by
\begin{linenomath*}
\begin{equation}
\label{newdef}
 \Gamma(E_\nu) = \Gamma_0 \bigg( \frac{E_\nu}{E_0} \bigg)^n ,
\end{equation}
\end{linenomath*}
\noindent where $n$ is the energy scaling power, $E_0$ is an arbitrary reference energy, and $\Gamma_0$ quantifies the decoherence strength at $E_0$. This power-law energy-dependence has also been assumed in previous experimental searches for neutrino decoherence~\cite{SKdeco,coloma2018decoherence,Gomes:2020muc}). The values of $n$ and $\Gamma_0$ are free parameters, and we take the approach of fixing $n$ and profiling over $\Gamma$.  Since the curvature of spacetime depends on energy density, it is natural to expect that the exponent $n$ would be positive.  Some models of quantum decoherence due to gravitational effects have suggested an $E_\nu^2$ energy scaling~\cite{ellis1997quantum2,ellis1997quantum,benatti1999non}.  The value of $\Gamma_0$ for a given model depends on $E_0$, which for convenience we have set to be close to the peak of the detected neutrino energy distribution at 1~TeV.  The value of $\tilde{\Gamma}_0$ for any other choice of $\tilde{E}_0$ can be obtained via $\Gamma_0=\tilde{\Gamma}_0 (E_0/\tilde{E}_0)^n$, and to compare with past experiments we present the final results at the reference energy of 1~GeV.  

A related quantity is the \textit{coherence length} $L_{\rm{coh}}$, defined as the distance at which the damping effects resulting from the loss of coherence reaches $1/e$ at a reference neutrino energy. In the models considered here, $L_{\rm{coh}}(E) = 1 / \Gamma(E)$ can be interpreted as the $\nu$VBH interaction mean free path~\cite{stuttard2020neutrino}, itself dependent on the VBH number density and $\nu$VBH interaction cross section.   A natural expectation for decoherence effects emerging from physics at the Planck scale can be obtained by setting $\Gamma(E_{P})\geq E_{P}$.  This corresponds to the assumption that one Planck-energy neutrino has a mean free path for interaction with a VBH of a Planck length or less, implying that  quantum coherence is effectively impossible at the Planck scale.  Our results  probe deep into this ``Planck scale naturalness'' region of parameter space for energy scaling factors $n\lesssim3$.

\section{Event Selection and Systematic Uncertainties}

This article presents an analysis that constrains $\nu$VBH decoherence models with IceCube data using 305,735 reconstructed up-going $\nu_\mu$ and $\bar{\nu}_\mu$ events in the energy range 0.5-10 TeV.  The sample was described in detail in Ref.~\cite{meowsprd}, and has been used by the IceCube collaboration for searches for eV-scale sterile neutrinos through matter-resonant oscillations~\cite{meows,abbasi2022search} and for neutrino-nucleus non-standard interactions~\cite{PhysRevLett.129.011804}.

The event selection provides a sample of track-like events produced by up-going muons traversing the detector.  In this energy regime the tracks tend to be well-reconstructed with energy resolution $\sigma_{\log _{10}(E_\mu)}\sim$0.3 and angular resolution $\sigma_{\cos\theta}$ varying between 0.005 and 0.015 as a function of energy. Since cosmic-ray muons are blocked from the upgoing flux by the Earth, the event selection is predicted to have a purity of $\gtrsim$99\% of muons generated by CC $\nu_\mu$ interactions below and within the IceCube detector.  In scenarios where high-energy $\nu_\tau$ appearance is enabled, such as decoherence models, the sample also contains a contribution from the 17\% of leptonic $\tau$ decays following $\nu_\tau$ CC interactions below or inside the detector volume.

A detailed discussion of the signal simulation and the suite of systematic uncertainties considered can be found in Ref. ~\cite{meowsprd} and is summarized briefly in the Supplementary Information.  Included in the uncertainty budget are parameters associated with the primary cosmic ray flux, hadronic interaction cross sections governing air-shower evolution, the astrophysical neutrino flux, detector performance parameters, ice properties, and neutrino interaction cross sections.  Each uncertainty contribution is treated as continuous nuisance parameter in a likelihood maximization analysis that compares the best-fit likelihood at various values of $\Gamma$ to establish a unified, frequentist confidence interval for each power index $n$ and decoherence model.  The likelihood test statistic follows Ref.~\cite{arguelles2019binned}  and is constructed to account for both data and Monte Carlo statistical uncertainties.

We constrain $\Gamma_0$ for $n$=0,1,2,3 using events binned logarithmically in reconstructed muon energy, $\log (E_{\text{reco}}^\mu)$ (13 bins, $E_{\text{reco}}^\mu \in [500\text{ GeV}\, , 9976\text{ GeV}]$), and uniformly in zenith angle (20 bins, $\cos(\theta_{\text{reco}}^\mu) \in [-1.0 \, , 0.0]$).  

\begin{figure}[t]
\centering
\includegraphics[width=.48\textwidth]{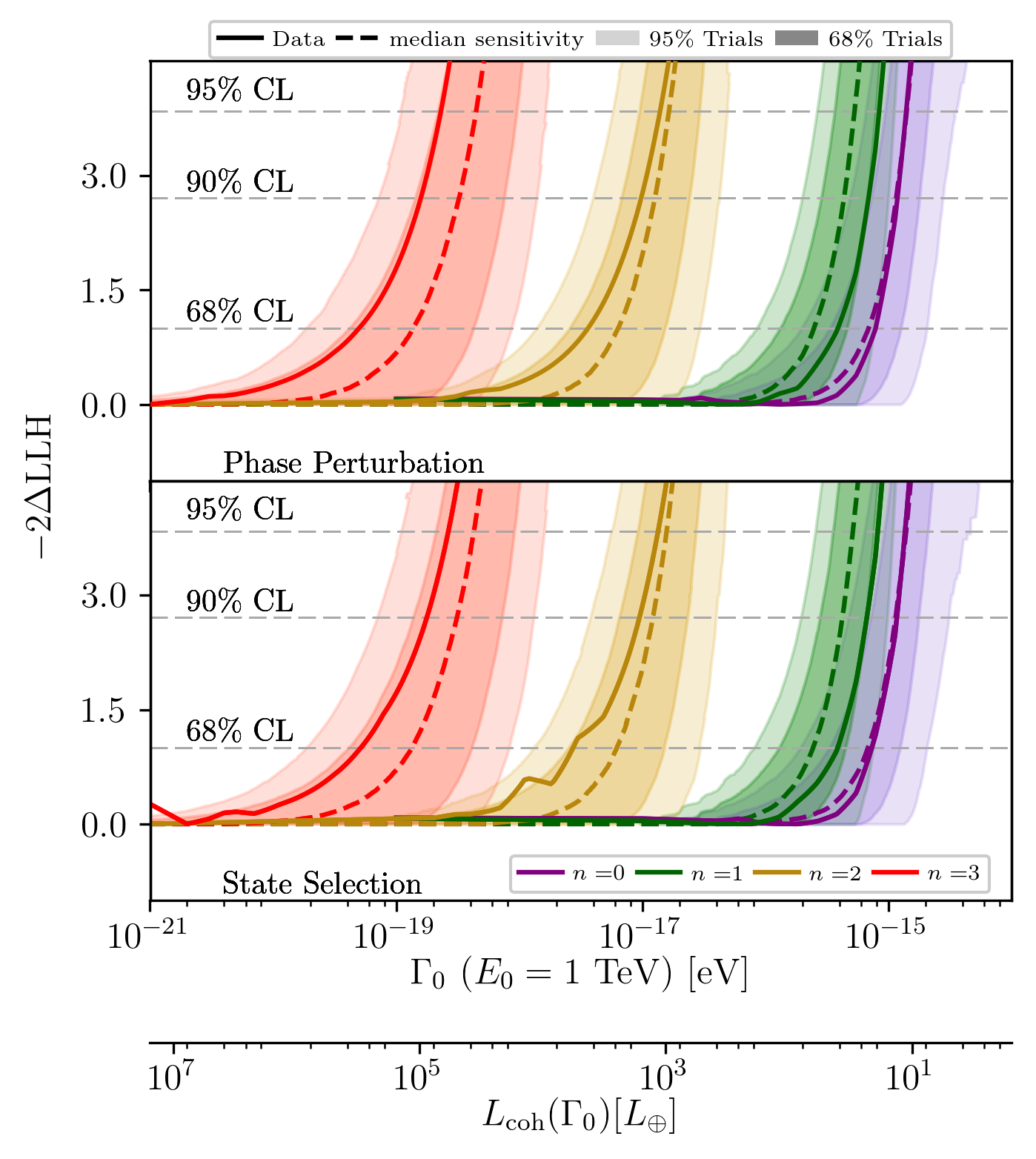}
\caption{\textit{\textbf{Test statistic distributions.}} 90\% CL sensitivity (dashed) and data (solid) -2$\Delta$LLH distributions for phase perturbation (top) and state selection (bottom). Frequentist regions for 95\% and 68\%  of 1,000 pseudoexperiment trials are also included (shaded region). The corresponding models of $n$ = 0, 1, 2, 3 are from right to left. Coherence distances at 1 TeV are shown relative to the Earth diameter $L_\oplus$.}
\label{fig:fig3}
\end{figure}

\begin{figure}[t]
\centering
\includegraphics[width=.41\textwidth]{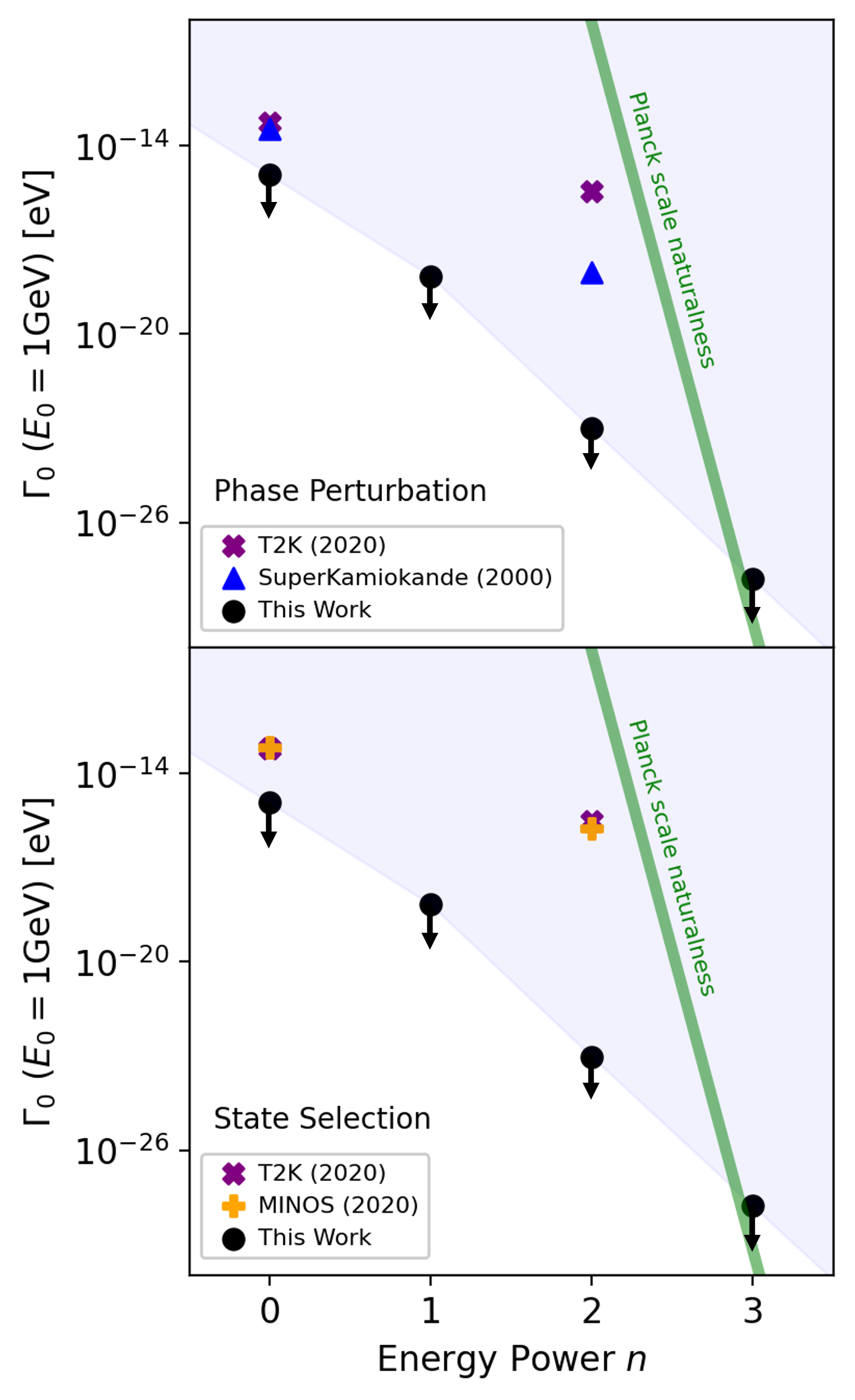}
\caption{\textit{\textbf{Comparison to previous results, from~\cite{Gomes:2020muc,SKdeco}.}} Top: Comparison of limits from both analyses on the phase perturbation interaction model to previous results. Bottom: Comparison of limits from both analyses on the state selection interaction model to previous results.}
\label{fig:fig4}
\end{figure}

\section{New Constraints on Anomalous Decoherence}

The analysis was developed blindly using simulated data and then applied to real data following a staged unblinding protocol developed for IceCube oscillation measurements.   Prior to unblinding, the median analysis sensitivity in the event of a null signal and its 68\% and 95\% envelopes were established using 1,000 Monte Carlo pseudoexperiments. The expected analysis performance in the event of an injected signal was also tested.  Signals beyond the 90\% contour were recovered exactly in un-fluctuated fits and with the expected level of accuracy when data fluctuations were included. 

A multi-stage blind fit procedure was followed, first checking energy and zenith pull distributions and then 1D histograms at the best-fit point, followed by nuisance parameter pulls, the joint [energy,zenith] distribution and pulls, and finally unblinding the full result.  The result is consistent with the null hypothesis for all decoherence models.  The p-value, defined as the fraction of simulated decoherence-free pseudoexperiments with test statistic larger than that observed in the data, is in the range 0.59-0.61 for each tested decoherence model and energy scaling power $n$.  The final upper limits on the decoherence parameter $\Gamma$  in all cases fall within the 68\% envelope of values expected if no decoherence is present, as shown in Fig.~\ref{fig:fig3}.  Feldman-Cousins ensemble tests were performed at the 90\% Confidence Level (CL) positions to check for proper coverage~\cite{feldman1998unified}.  The 90\% CL locations from Wilks' theorem were found to be slightly weaker than the Feldman Cousins values, with a maximal deviation of 28.2\% in the value of $\Gamma$ at 90\% CL. This difference is imperceptible on the logarithmic $\Gamma$ axes of Figs.~\ref{fig:fig3} and ~\ref{fig:fig4}.  The 90\% confidence limits from this analysis $\Gamma_{90}$($E_0$=1~TeV) for the State Selection (SS) and Phase Perturbation (PS) scenarios are tabulated in Tbl~\ref{table:ssup}. For reference, Fig.~\ref{fig:fig3} also lists the corresponding coherence lengths for a 1~TeV neutrino.

\begin{table}[t]
\centering
\begin{tabular}{|c|c|c|} 
   \hline

   \headercell{$n$} & \headercell{Phase Perturbation~$\Gamma_{90}$} & \headercell{State Selection~$\Gamma_{90}$}\\
 \hline
     0 & $1.18\cdot 10^{-15}$~eV&$1.17\cdot 10^{-15}$~eV \\
     1 & $6.89\cdot 10^{-16}$~eV&$6.67\cdot 10^{-16}$~eV \\
     2 & $9.80\cdot 10^{-18}$~eV&$9.48\cdot 10^{-18}$~eV \\
     3 & $1.58\cdot 10^{-19}$~eV&$1.77\cdot 10^{-19}$~eV \\
    \hline

\end{tabular}
\vspace{0.5cm}
\caption{The 90\% CL upper limits on $\Gamma_{90}$  for each $n$ in $E_0$=1~TeV in the state selection (SS) and phase perturbation (PS) models.}
\label{table:ssup}
\end{table}

We thus report strong new constraints on the magnitude of anomalous decoherence from quantum gravity as neutrinos oscillate across the Earth.  Using flavor structures motivated by neutrino interactions with space time foam through $\nu$VBH interactions and four power law models,  90\% CL limits have been obtained on the parameter $\Gamma_0$ with pivot energy $E_0$=1~TeV.  To facilitate comparison to previous studies, our results can be mapped to the pivot energy 1~GeV that has tended to be favored by previous works.  A comparison to past data is shown in Fig.~\ref{fig:fig4}.  For all tested power law indices and both tested flavor models, the results presented in this paper provide the world's strongest limits.   In all cases with $n<3$ the new limits significantly surpass the natural Planck scale benchmark.

The energy-independent ($n=0$) scenario has been explored previously for T2K \cite{Gomes:2020muc}, Super-Kamiokande \cite{SKdeco}, and MINOS \cite{Gomes:2020muc}.  The IceCube limits extend beyond past measurements by a factor of around 30 for the state selection model and 50 for phase perturbations, thanks to the larger sample size of the IceCube atmospheric neutrino data set.   

The substantially increased energy range of the IceCube atmospheric neutrino sample leads to a far more dramatic improvement in sensitivity for models where the decoherence strength depends on energy with a positive exponent. Past results have been obtained for the gravity-motivated $n=2$ model ~\cite{ellis1997quantum2,ellis1997quantum,benatti1999non} by the aforementioned three experiments, which primarily collect neutrinos at around three orders of magnitude lower energy than the peak of the IceCube sample.   We report improvements by six orders of magnitude in the phase perturbation model, and eight orders of magnitude in the state selection model.  Since quantum gravitational effects are anticipated to scale positively with energy density, the limits presented in this paper represent a major increase in sensitivity to anomalous decoherence from quantum gravity in the neutrino sector.

\section*{acknowledgements}
USA {\textendash} U.S. National Science Foundation-Office of Polar Programs,
U.S. National Science Foundation-Physics Division,
U.S. National Science Foundation-EPSCoR,
U.S. National Science Foundation-Office of Advanced Cyberinfrastructure,
Wisconsin Alumni Research Foundation,
Center for High Throughput Computing (CHTC) at the University of Wisconsin{\textendash}Madison,
Open Science Grid (OSG),
Partnership to Advance Throughput Computing (PATh),
Advanced Cyberinfrastructure Coordination Ecosystem: Services {\&} Support (ACCESS),
Frontera computing project at the Texas Advanced Computing Center,
U.S. Department of Energy-National Energy Research Scientific Computing Center,
Particle astrophysics research computing center at the University of Maryland,
Institute for Cyber-Enabled Research at Michigan State University,
Astroparticle physics computational facility at Marquette University,
NVIDIA Corporation,
and Google Cloud Platform;
Belgium {\textendash} Funds for Scientific Research (FRS-FNRS and FWO),
FWO Odysseus and Big Science programmes,
and Belgian Federal Science Policy Office (Belspo);
Germany {\textendash} Bundesministerium f{\"u}r Bildung und Forschung (BMBF),
Deutsche Forschungsgemeinschaft (DFG),
Helmholtz Alliance for Astroparticle Physics (HAP),
Initiative and Networking Fund of the Helmholtz Association,
Deutsches Elektronen Synchrotron (DESY),
and High Performance Computing cluster of the RWTH Aachen;
Sweden {\textendash} Swedish Research Council,
Swedish Polar Research Secretariat,
Swedish National Infrastructure for Computing (SNIC),
and Knut and Alice Wallenberg Foundation;
European Union {\textendash} EGI Advanced Computing for research;
Australia {\textendash} Australian Research Council;
Canada {\textendash} Natural Sciences and Engineering Research Council of Canada,
Calcul Qu{\'e}bec, Compute Ontario, Canada Foundation for Innovation, WestGrid, and Digital Research Alliance of Canada;
Denmark {\textendash} Villum Fonden, Carlsberg Foundation, and European Commission;
New Zealand {\textendash} Marsden Fund;
Japan {\textendash} Japan Society for Promotion of Science (JSPS)
and Institute for Global Prominent Research (IGPR) of Chiba University;
Korea {\textendash} National Research Foundation of Korea (NRF);
Switzerland {\textendash} Swiss National Science Foundation (SNSF).

\bibliography{sn-bibliography}


\section{Supplementary Information}

\subsection{Simulation methods and systematic uncertainties}
To predict the expected rate of upgoing tracks selected as a function of zenith and energy, a three-flavor prediction of neutrino flux emerging from cosmic-ray air showers is first made using the {\tt MCEq} software package~\cite{Fedynitch:2015zma}.

\begin{table}[b!]
\begin{adjustbox}{width=\columnwidth,center}
\begin{tabular}{|l|c|c|c|} 
\hline
\multicolumn{1}{|c|}{} & \multicolumn{1}{|c|}{} & \multicolumn{2}{|c|}{\textbf{Value at Best Fit ($n=0$)}}\\

\multicolumn{1}{|c|}{\textbf{Parameter}} & \multicolumn{1}{|c|}{\textbf{Prior}} & \multicolumn{1}{|c|}{\textbf{Phase}} & \multicolumn{1}{|c|}{\textbf{State}}\\ 
\multicolumn{1}{|c|}{} & \multicolumn{1}{|c|}{} & \multicolumn{1}{|c|}{\textbf{Perturbation}} & \multicolumn{1}{|c|}{\textbf{Selection}}\\ 
\hline 
\multicolumn{4}{|c|}{\multirow{2}{*}{\textbf{Detector Parameters}}} \\  
\multicolumn{4}{|c|}{}\\ \hline
DOM Efficiency & 0.97 $\pm$ 0.10 & 0.96& 0.96 \\ 
Bulk Ice Gradient 0 & 0.0$\pm$ $1.0^*$&-0.08&-0.07\\
Bulk Ice Gradient 1 & 0.0$\pm$ $1.0^*$&0.58&0.60\\
Forward Hole Ice ($p_2$)&-1.0$\pm$10.0&-3.32&-3.30\\ \hline
\multicolumn{4}{|c|}{\multirow{2}{*}{\textbf{Conventional Flux Parameters}}}\\ 
\multicolumn{4}{|c|}{}\\ \hline
Normalization$(\Phi_{conv.})$ & 1.0$\pm$0.4&1.10&1.10\\
Spectral Shift ($\Delta_{conv.}$) &0.00$\pm$0.03&0.07&0.07\\
Atmospheric Density &0.0$\pm$1.0 &-0.10&-0.11\\
Barr WM &0.00$\pm$0.40&-0.01&-0.00\\ 
Barr WP &0.00$\pm$0.40&0.01&0.01\\ 
Barr YM &0.00$\pm$0.30&-0.06&-0.06\\ 
Barr YP &0.00$\pm$0.30&-0.04&-0.04\\ 
Barr ZM &0.00$\pm$0.12&-0.01&-0.01\\ 
Barr ZP &0.00$\pm$0.12& -0.02&-0.02\\ \hline
\multicolumn{4}{|c|}{\multirow{2}{*}{\textbf{Astrophysical Flux Parameters}}}\\ 
\multicolumn{4}{|c|}{}\\ \hline
Normalization$(\Phi_{astro.})$ &0.79$\pm$ $0.36^*$ &0.84&0.84\\
Spectral Shift $(\Delta_{astro.})$ & 0.00$\pm$ $0.36^*$ &-0.02&-0.01\\ \hline
\multicolumn{4}{|c|}{\multirow{2}{*}{\textbf{Cross sections}}}\\ 
\multicolumn{4}{|c|}{}\\ \hline
Cross section ($\sigma_{\nu_\mu}$) &1.00$\pm$0.03&1.00&1.00 \\
Cross section ($\sigma_{\overline{\nu}_\mu}$) &1.00$\pm$0.075&1.01&1.01\\
Kaon re-interaction ($\sigma_{KA})$& 0.0$\pm$1.0&-0.16&-0.14\\ \hline
\end{tabular}
\end{adjustbox}
\vspace{0.5cm}
\caption{\textbf{Summary of nuisance parameters used in the analysis}. Each row specifies the constraint used in the frequentist analysis for each physics or nuisance parameter. All priors are one dimensional Gaussian functions, except in the case of the bulk ice and astrophysics flux parameters (marked with an asterisk) where a correlated prior is employed. The value of the nuisance parameters at best-fit point for $n=0$ are given for both state selection and phase perturbation interaction modes.  A fully detailed description of these parameters and their technical implementations can be found in Ref~\cite{meowsprd}.
}
\label{table:systs}
\end{table}

The simulation of neutrino flux with MCEq employs the {\tt SIBYLL} cosmic-ray interaction model~\cite{ahn2009cosmic,riehn2020hadronic}, varied within systematic uncertainties as described by Barr {\em et. al}~\cite{barr2006uncertainties}.  Neutrinos from the decays of heavy hadrons are also incorporated, according to the model of BERSS~\cite{bhattacharya2015perturbative}. The uncertainty deriving from production height dependence is insignificant for these energies. Uncertainties in the re-interaction cross section of kaons as they travel through the atmosphere are relevant and are included. 

\begin{figure}[t]
    \centering
    \includegraphics[width=.48\textwidth]{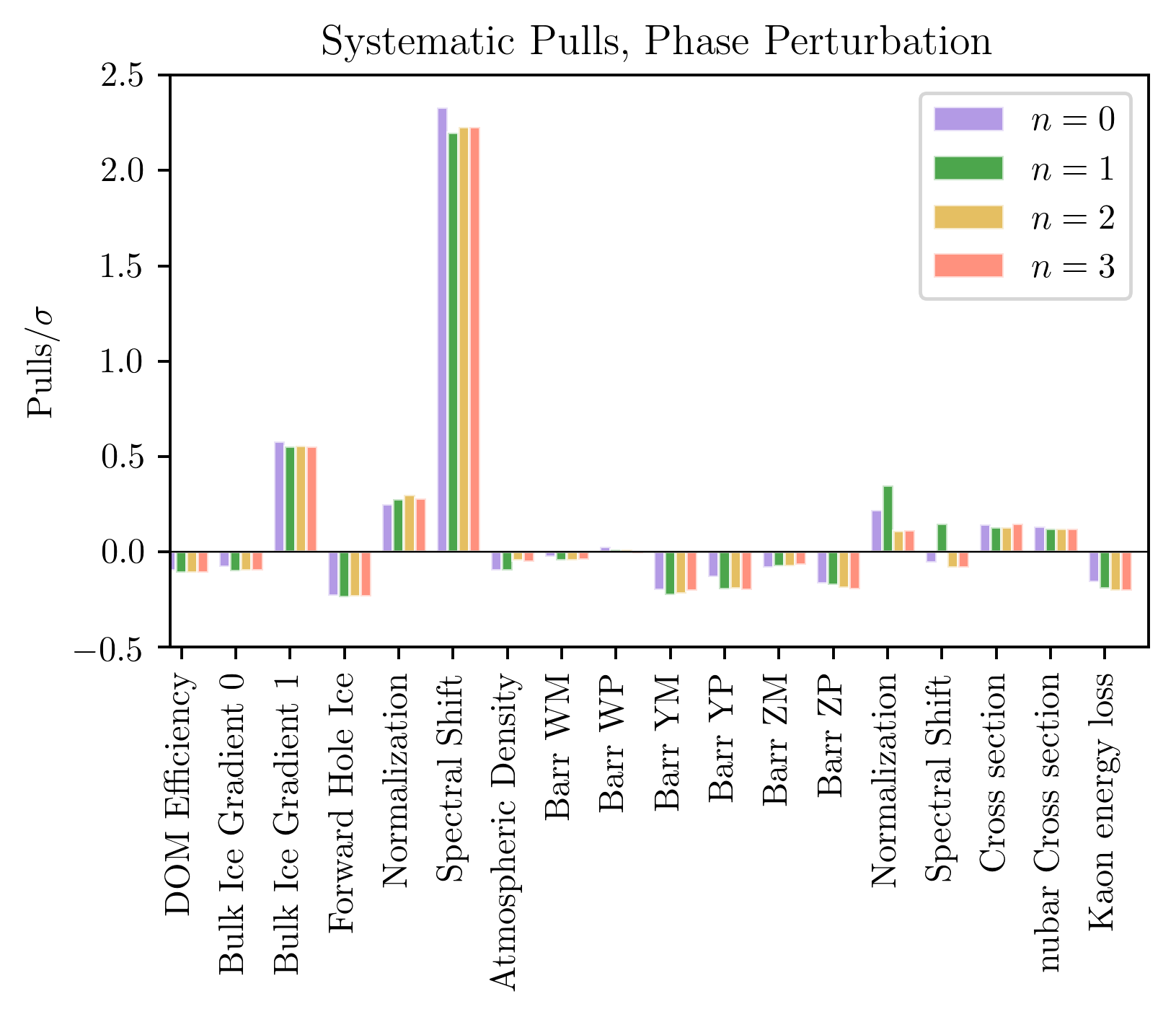}
    \caption{\textbf{Systematic Pulls for Phase Perturbation}}
    \label{fig:i}
\end{figure}
\begin{figure}[t]
    \centering
    \includegraphics[width=.48\textwidth]{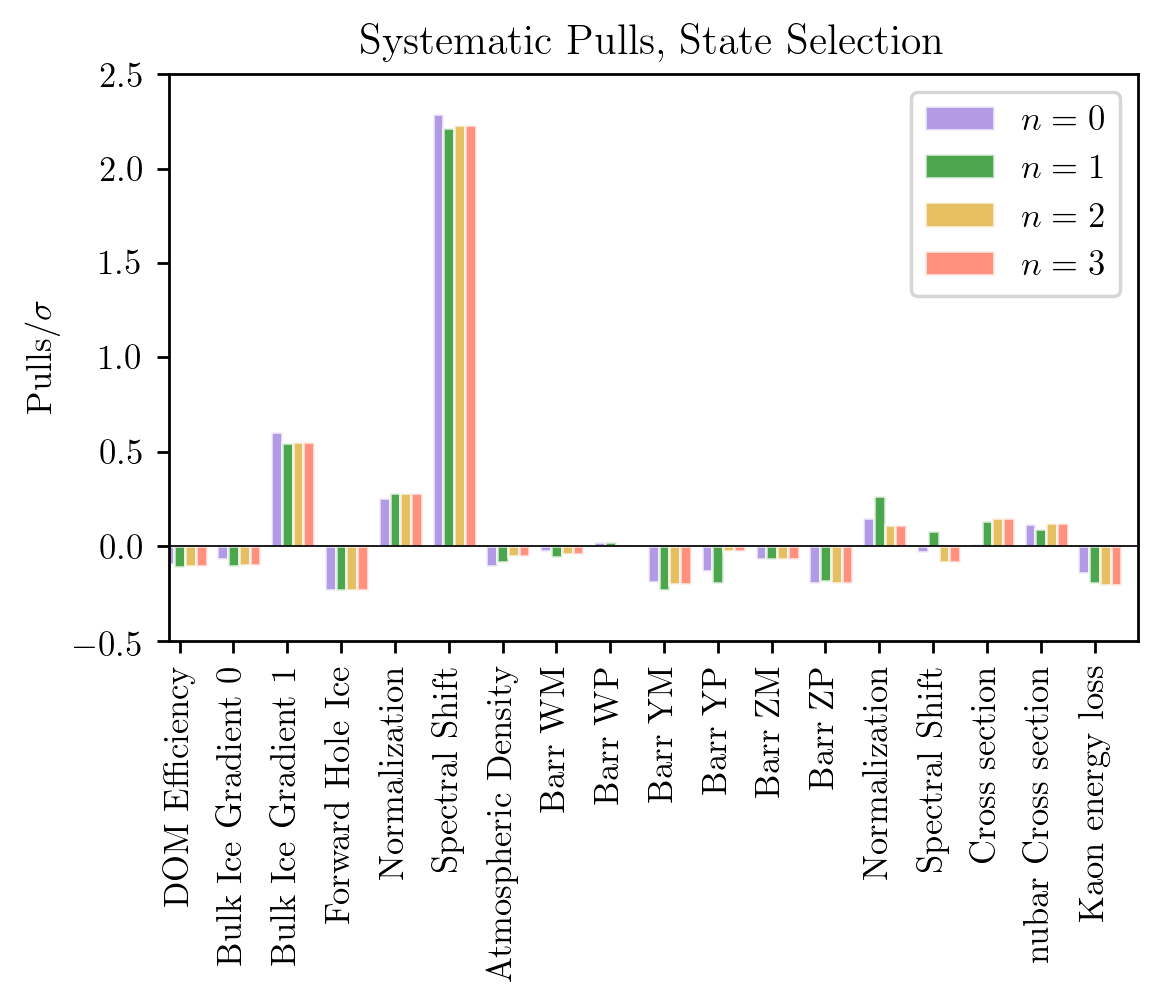}
    \caption{\textbf{Systematic Pulls for State Selection}}
    \label{fig:ii}
\end{figure}

The astrophysical neutrino flux is simulated and propagated using parameters informed by existing IceCube measurements~\cite{Aartsen:2020aqd}, but with a wide systematic uncertainty on normalization and spectral index.  The central model is taken to be a single unbroken power law in neutrino energy with spectral index of 2.5.  Variation of the astrophysical flux as a result of decoherence was tested and found to lead to negligible effects within the analysis sample. Each flux is propagated under the three-flavor oscillation formalism including decoherence, absorption and tau regeneration with the {\tt nuSQuIDS} package~\cite{arguelles2022nusquids} to provide an energy, zenith and flavor dependent prediction at the detector for each oscillation hypothesis. The detector response to each flux is calculated by applying the reweighting protocol described in Ref.~\cite{leptonsoftware} to a large Monte Carlo event ensemble to predict the final energy spectrum, which is used as input to calculate the analysis likelihood test statistic. 

All systematic uncertainties are implemented as continuous nuisance parameters that are fitted through a profile log-likelihood analysis.  The test statistic incorporates both accurate treatment of low-population bins and also the effects of finite Monte Carlo sample size, and is discussed in Ref. ~\cite{arguelles2019binned}. The dominant sources of systematic uncertainty in this analysis are the detector performance uncertainties, which include the photon detection efficiency of the DOMs~\cite{abbasi2010calibration}, the properties of the refrozen ice in the vicinity of the detector strings~\cite{rongen2016measuring}, and the properties of the bulk ice in the array~\cite{aartsen2013measurement}.    Depth-dependent uncertainties on optical absorption and scattering in the bulk glacial ice are accounted for following the method described in~\cite{aartsen2019efficient}.  An improved ice model that incorporates crystal birefringence as a mechanism to explain the observed anisotropic light propagation in the glacier~\cite{abbasi2022situ} was also tested within this analysis as a post-unblinding check, and was not observed to affect the results significantly.

Sub-leading systematic uncertainties associated with the primary cosmic-ray flux, evolution of atmospheric neutrino air showers, atmospheric density effects, astrophysical neutrino flux normalization and spectral shape, and neutrino cross section, are also included. The full list of systematic uncertainties, their Gaussian priors, and the best-fit values in one example decoherence scenario is shown in Tbl.~\ref{table:systs}. The full distribution of pulls for each model point is shown in Fig~\ref{fig:ii}. The largest deviation on any systematic uncertainty is the cosmic-ray spectral index which pulls to $2.3\sigma$ at its best-fit point.  No other parameter deviates further than $1\sigma$ from the center of its prior.  This behaviour is qualitatively similar to what has been observed with previous IceCube analyses that have employed this sample to search for BSM oscillation physics~\cite{PhysRevLett.129.011804,meows,abbasi2022search}.

\end{document}